# Autonomous Kinetic Modeling of Biomass Pyrolysis using Chemical Reaction Neural Networks

Weiqi Ji [a,1], Franz Richter [b,1], Michael J. Gollner [b], Sili Deng [a,*]

[a] Department of Mechanical Engineering, Massachusetts Institute of Technology, Cambridge, MA 02139, USA.

[b] Department of Mechanical Engineering, University of California, Berkeley, CA 94701, USA.

[1] Weiqi Ji and Franz Richter contribute equally to this work.

[*] Corresponding author: Sili Deng (silideng@mit.edu).

## Abstract

Modeling the burning processes of biomass such as wood, grass, and crops is crucial for the modeling and prediction of wildland and urban fire behavior. Despite its importance, the burning of solid fuels remains poorly understood, which can be partly attributed to the unknown chemical kinetics of most solid fuels. Most available kinetic models were built upon expert knowledge, which requires chemical insights and years of experience. This work presents a framework for autonomously discovering biomass pyrolysis kinetic models from thermogravimetric analyzer (TGA) experimental data using the recently developed chemical reaction neural networks (CRNN). The approach incorporated the CRNN model into the framework of neural ordinary differential equations to predict the residual mass in TGA data. In addition to the flexibility of neural-network-based models, the learned CRNN model is interpretable, by incorporating the fundamental physics laws, such as the law of mass action and Arrhenius law, into the neural network structure. The learned CRNN model can then be translated into the classical forms of biomass chemical kinetic models, which facilitates the extraction of chemical insights and the integration of the kinetic model into large-scale fire simulations. We demonstrated the effectiveness of the framework in predicting the pyrolysis and oxidation of cellulose. This successful demonstration opens the possibility of rapid and autonomous chemical kinetic modeling of solid fuels, such as wildfire fuels and industrial polymers.

# 1. Introduction

Pyrolysis is a key driver for wildland and urban fires alike. For example, in wildfires, pyrolysis produces the gaseous fuel to sustain the spread of flames, while in timber buildings, it controls the decay of a building's structural integrity during a fire. Despite its importance, pyrolysis remains poorly understood [1] which can be partially attributed to the unknown chemical kinetics of most solid fuels. This lack of knowledge stems from the fact that multi-step chemical kinetic models are built from the chemical insight of experts acquired over years, which makes the process slow and cumbersome. This knowledge is unavailable for biomass fuels, which partly hinders the advancement of modeling wildfires [2], biofuels [3], and bio-based construction materials [4].

Neural network models have been employed for combustion kinetic modeling for computation acceleration by approximating a given complex model with computationally cheap neural network models [5–9], and for learning combustion models from experimental data when there is no available model [10,11]. While black-box neural network models have shown success in fitting the experimental data, it is often desirable that the learned model can be interpretable such that the model can also elucidate the reaction pathways and thus provide chemical insights. In addition, a model that can be interpreted and consistent with fundamental physical laws are more likely to be able to extrapolate beyond the range of thermodynamic conditions where the model is learned.

Therefore, this work aims at developing an interpretable neural network biomass pyrolysis modeling approach that can not only fit the experimental data but also elucidate the reaction pathways and kinetic parameters. We adopt the recently developed framework of Chemical Reaction Neural Networks (CRNN), which achieves interpretability by incorporating the fundamental physical laws such as the law of mass action and the Arrhenius law into the structure of the neural networks. Both the reaction pathways (stoichiometric coefficients) and kinetic rate constants are treated as learnable parameters such that prior knowledge of reaction pathways are not required. We augment the CRNN with neural ordinary differential equations to facilitate the training of CRNN from TGA datasets. Since the number of model parameters can easily go beyond one hundred, heuristic optimization methods, such as genetic algorithms, particle swarms, and evolution algorithms, which are widely adopted in previous biomass kinetic modeling frameworks [12], are not suitable. We thus adopt stochastic gradient descent (SGD) optimization to train the CRNN model given the success of SGD in optimizing high-dimensional nonlinear deep neural network models [13].

We then demonstrate the approach in modeling cellulose pyrolysis. The cellulose model is a building block for more complex biomass fuels and has been extensively studied in the literature, which allows us to compare the learned model with existing models derived with expert knowledge. Since the proposed approach does not require knowledge of the target fuel, the framework could be applied to a variety of fuels, such as bio-derived and engineered polymers. Therefore, the approach has the potential to transform the way we study the pyrolysis kinetics for biomass production and fire safety modeling.

The structure of this work is as follows: we shall first review classical modeling and experimental approaches for studying the chemical kinetics of solid fuels, and then introduce the framework of CRNN and the learning results from experimental data compiled from literature; finally, we discuss the opportunities of coupling CRNN with gas-phase kinetics and diffusion processes.

## 2. Literature Review on Kinetic Modeling of Solid Fuels

### 2.1. Classic Chemical Kinetics of Solid Fuels

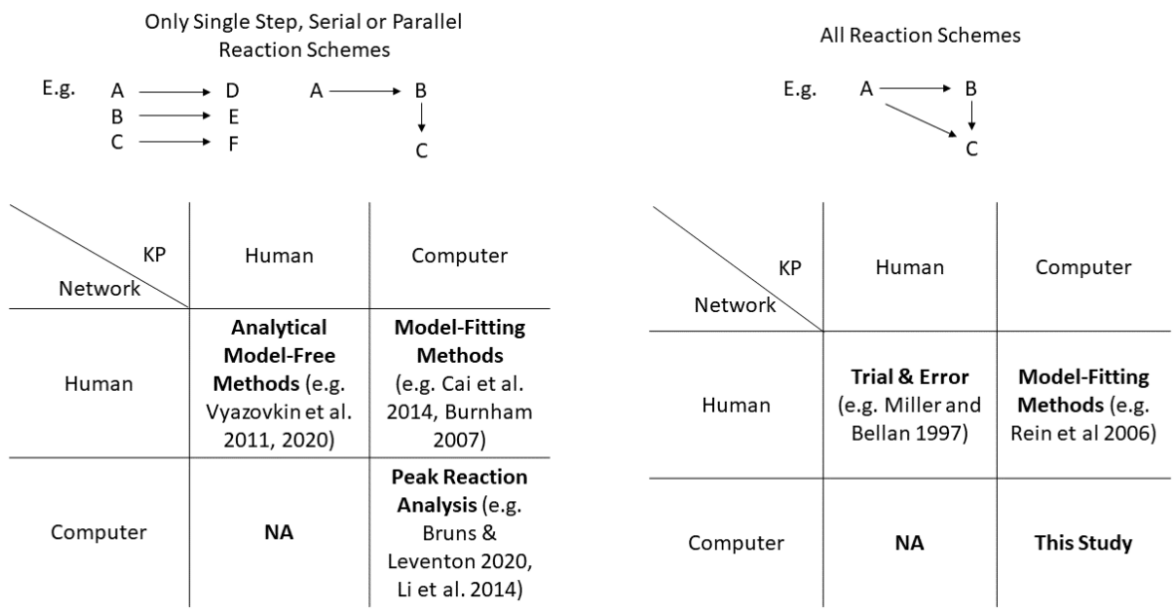

Figure **1**. Framework of the current literature on determining kinetic parameters (KP) and reactions schemes (network). The paper referenced are [14–21].

This discussion will restrict itself to the derivation of reduced chemical kinetic models that can be utilized by existing combustion and fire codes at a large scale. The importance of chemical kinetic models for large-scale fire modeling via scaling has been described by Torero [1,22,23]. The standard experimental method to derive reduced kinetic models is a Thermal Gravimetric Analyzer (TGA), which is a small furnace able to record the mass-temperature histories of small samples. Usually, a sample of a few milligrams is heated at a constant rate in a TGA while one measures simultaneously the mass and temperature of the sample. The sample is small to ensure that the sample has the same temperature as the surrounding fluid—this ensures that the actual rate of heating of the sample is equal to the desired rate—and that the lumped capacitance assumption holds. The latter ensures that the sample is at a uniform temperature. These mass-temperature histories are then presented as normalized mass-temperature histories (where the mass is normalized to the initial mass) or conversion- temperature histories (where the mass is normalized to the initial and final mass). Commonly, the derivative of the analyzed mass or conversion history is also taken.

Figure **1** provides an overview of the available types of methods to derive kinetic models from TGA experiments and whether they rely on humans or computers to determine kinetic parameters and reaction schemes (networks). Classical reduced chemical kinetic modeling of solid fuels can be split into two methods: model-free and model-fitting. They concern themselves with only single, parallel, or serial reaction schemes. The model-free approach is favored by the thermogravimetric community, which relies on conducting experiments in a TGA at constant heating rates and calculating the derivative of the conversion- temperature history. There are then a variety of methods that utilize the peak and shape of the resulting curve to develop kinetic parameters and a mathematical formulation of the rate constant and law of mass action [14]. The advantages of this method are that it is established and takes a relatively short time to conduct with the best practice available [14]. The disadvantage is that all model-free methods are restricted to a one-step reaction, they predict fixed final product yields, and the kinetic parameters vary with conversion. In practice, however, final product yields vary, large-scale codes require invariant kinetic parameters (no change with conversion), and many fuels show behavior only captured by multiple reaction steps.  The interpretation and translation of the results of model-free methods into practical usefulness, therefore, requires expert chemical knowledge and experience. Most model-free methods also suffer from errors due to using only one heating rate or only relying on the peak in the mass-loss rate curve [24]. The model-fitting approach is favored by researchers in the biomass and fire science community. In this approach, a reaction scheme is derived based on existing chemical knowledge. The rate constant and the law of mass action are formulated, similar to gas-phase

kinetics, in a standardized way [25]. The kinetic parameters are found by fitting the chemical kinetic model to several experiments. This approach has the advantage of predicted variable product yields, reduced compensation errors due to the use of larger sets of experiments, being directly incorporable into large-scale fire and combustion models, the ability to incorporate multiple reactions, and making use of chemical insight from different sources such as other experiments or atomistic modelling. Over the past two decades, this approach has yielded more advances than the model-free method through models such as the Ranzi scheme [26]. The disadvantage of this approach is that developing a reaction scheme requires expert knowledge. In other words, model-free methods are easy to use but hard to interpret, while model-fitting methods are hard to use (well) but easy to interpret.

Over the last five to ten years, the communities have moved in different directions. The thermogravimetric community as well as the fire science community have worked on combining the model-free and the model-fitting methods [15,18,27–30]. Generally, the idea is to fit individual reactions to the individual peaks in the mass loss rate curve of a TGA experiment to find the number of reactions and an initial guess for the kinetic parameters. The parameters are then optimized to all heating rates to improve them. An example is Peak Reaction Analysis (Figure 1). This approach enables the researchers to combine most of the advantages of both methods, but it limits the reaction schemes to be either all sequential or parallel reactions. As a result, these new approaches still suffer from the inability to predict variable product yields that require competing reactions. They also fail to take into account isothermal experiments, which are advised to be reproduced as well [24]. On the other hand, the biomass and combustion community have moved towards mechanistic models to first unravel the intrinsic kinetics of biomass and derive reduced kinetic approaches from there. This approach has provided chemical insight, but no significant step in reduced kinetic modelling [31]. It was argued that detailed chemical kinetics is presently not suitable for problems in the fire community as well as elsewhere [32]. Over the last few years, the biomass community has also applied machine learning with success in some specific problems [31], but these findings are unfortunately not extendable to wildfires.

In short, most current methods require significant expertise to derive a complex multi-step kinetic model. As a result, kinetic models are only available for a small number of fuels. There is a need to automate the development of chemical kinetic models for solids to create models for a wide collection of solid fuels.

## 2.2. Cellulose Kinetic Models and Experiments

The chemical kinetics of cellulose is established and has recently been well-reviewed [31,33–35]. One can divide all models into three levels of complexity. The first level applies to applications where the details of chemical kinetics are unimportant. It was found that a single-step reaction with a sigmoidal character is appropriate [35]. The kinetic parameters ($A \sim 10^{15}\ s^{-1}, E_a \sim 47 - 52$ kcal/mol) are also consistent with atomistic calculations. In the second level, chemical kinetics is important, but the chemical kinetic model is a sub-model of a larger calculation. One example is the reduced kinetic models in the combustion and fire science community. Here, variations of the Broido-Shafizadeh model (around three reactions) were found to be most appropriate. They allow a nearly 20 % improvement in accuracy, the ability to predict a variable char yield with heating rate, and consistency with most experimental observations and atomistic calculations [33]. The last level of complexity is the mechanistic models that are used to gain chemical insight. They are derived from atomistic calculations and can contain over 100 reactions and species. Their main use is to optimize the yield of various high-value oils and gases from biomass. In the fire and combustion community, they serve as additional evidence for choosing a reduced kinetic model and a set of kinetic parameters. Cellulose, therefore, represents a good test case as its kinetics is well established and studied at all three levels of complexity.

## 3. Methodology

The code and experimental datasets presented in this work can be publicly accessed on GitHub at https://github.com/DENG-MIT/Biomass.jl.

### 3.1. Chemical Reaction Neural Network Method for Biomass Pyrolysis

Details of the Chemical Reaction Neural Network (CRNN) can be found in [36]. Here, we shall briefly present the formula of CRNN and incorporate constraints specific to biomass pyrolysis into the framework. Without loss of generality, we derive the CRNN for a system consisting of two intermediate species $[S_2, S_3]$. Together with cellulose, oxygen and the volatile matters, the system is comprised of five species, $[Cellu, S_2, S_3, O_2, Vola]$. Take the following reaction as an example:

$$v'_1 Cellu + v'_2 S_2 + v'_o O_2 \rightarrow v''_3 S_3 + v''_g Vola. \qquad (1)$$

With the law of mass action and Arrhenius law, we could write down the reaction rate and re-arrange it as

$$r = [Cellu]^{v'_1}[S_2]^{v'_2}[O_2]^{v'_o} AT^b exp(-E_a/RT) \qquad (2)$$

$$r = \exp\left(v_1' \ln[Cellu] + v_2' \ln[S_2] + v_O' \ln[O_2] + \ln A + b \ln T - E_a/RT\right), \qquad (3)$$

where $v_i'$ and $v_i''$ correspond to the stoichiometric coefficients of the reactants and products, respectively. Here, we assume that the reaction orders are equal to the stoichiometric coefficients. The rate constants consist of three parameters, and they are the pre-factor $A$, non-exponential temperature dependence factor $b$, and the activation energy $E_a$. We can further write down the production rate of each species as

$$\begin{aligned}
\frac{d[Cellu]}{dt} &= [\dot{Cellu}] = -v_1' r \\
\frac{d[S_2]}{dt} &= [\dot{S_2}] = -v_2' r \\
\frac{d[S_3]}{dt} &= [\dot{S_3}] = v_3'' r \\
\frac{d[Vola]}{dt} &= [\dot{Vola}] = v_g'' r.
\end{aligned} \qquad (4)$$

Recall the formula of a neuron in the neural network, $y = \sigma(\mathbf{w}\mathbf{x} + b)$, in which $\mathbf{x}$ is the input to the neuron, $y$ is the output, $\mathbf{w}$ are the weights, $b$ is the bias, and $\sigma(\cdot)$ is the nonlinear activation function. We can represent the reaction in Eq. **1** as a neuron with the activation function of an exponential function, as shown in Fig. **2a**. The inputs are the species concentrations in the logarithmic scale and the temperature, and the outputs are the production rates of all species, except for oxygen. The weights in the input layer correspond to the reaction orders, the Arrhenius parameters $b$ and $E_a$. The bias corresponds to the pre-factor $A$ in the logarithmic scale. The output weights correspond to the stoichiometric coefficients.

In general, the biomass pyrolysis process involves multiple steps which can be represented by stacking those single neurons to form a neural network with one hidden layer, as shown in Fig. **2b**. The number of hidden nodes is equal to the number of reactions. To accommodate kinetic models that are consistent with our common understanding of the biomass pyrolysis process, we incorporate the following constraints to the CRNN model for cellulose kinetic modeling.

(1) Cellulose (Cellu) is only present in the reactants.
(2) Volatile (Vola) gas is only present in the products.
(3) Mass stoichiometry is balanced.
(4) Activation energy $E_a$ is within the range of [0, 300] kcal/mol, and $\ln A$ is within [-23, 50].

In addition, the reaction orders are assumed to be equal to the stoichiometric coefficients for the reactants, so that the input and output weights are shared. It should be noted that the reaction orders are zero for products. Weight sharing usually helps reduce the cost of optimization, regularize the model, and prevent overfitting. In addition, bounding the stoichiometric coefficients and the reaction orders can prevent the model from being too stiff, blowing up the training, as it's challenging to train very stiff neural ordinary differential equations.

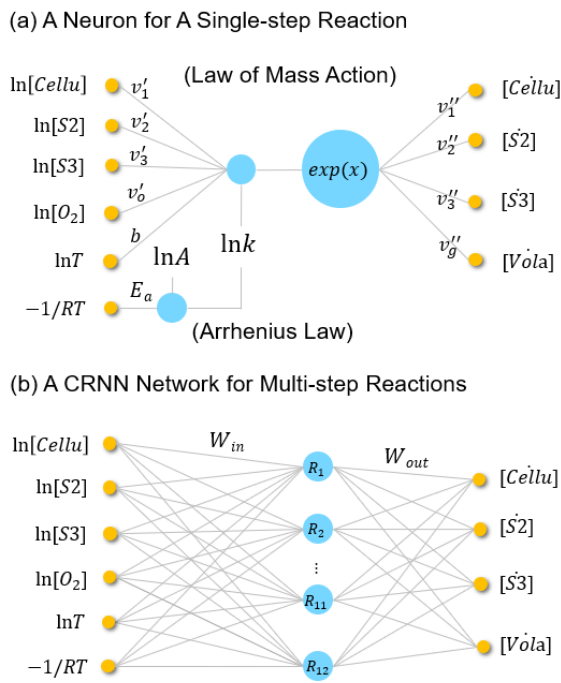

**Figure 2.** Schematic of the CRNN illustrated for a reaction system with four species and four reaction steps. (a) A neuron is designed based on the mass action law and the Arrhenius law, such that the learned neural network can be interpreted and translated into the traditional chemical reaction network. (b) Neurons are stacked into one hidden layer to formulate a CRNN for multi-step reactions.

As the weights and biases in the CRNN are physically interpretable, the CRNN is essentially the digital twin of the classical chemical reaction network. The inference of the reaction network can then be accomplished by training the CRNN with experimental measurements. Considering a general chemical reaction system where the vector of species concentration $Y$ evolves with time, we try to discover a CRNN that satisfies

$$\dot{Y} = CRNN(Y). \tag{5}$$

The CRNN can be trained from the concentration and production rate data pair of $\{Y, \dot{Y}\}$. However, in practice, we are usually unable to measure all of the concentrations of intermediate species and their derivatives. Furthermore, it is even challenging to know how many intermediate species there are and what the intermediate species are. On the other hand, we have the time-series of residual mass from TGA, which corresponds to the total mass of all species in the condensed phase.

Therefore, we augment CRNN with neural ordinary differential equations [37] to predict the residual mass and backpropagate the error into the CRNN model parameters. Specifically, an ordinary differential equation (ODE) system can be formulated with Eq. **5** and numerically solved in an ODE integrator by providing initial compositions and experimental conditions of the initial temperature $T_0$, heating rate $\beta$, and oxygen concentrations $[O_2]$. The solution is denoted as $Y^{CRNN}(t)$ in Eq. **6**, in which $Y_0$ is a vector containing the initial concentrations. The predicted residual mass can be readily derived from the species concentrations. We then define the loss function as the difference between the measured and predicted time series of residual mass. In the present work, the mean absolute error (MAE) is utilized as the loss metric in Eq. **6**.

$$Y^{CRNN}(t) = \text{ODESolve}(\text{CRNN}(Y), Y_0; T_0, \beta, [O_2]) \tag{6}$$

$$\text{Loss} = \text{MAE}\left(m_{res}^{CRNN}(t), m_{res}^{data}(t)\right) + \text{MAE}\left(m_{gas}^{CRNN}(t), m_{gas}^{data}(t)\right) \tag{7}$$

$m_{res}$ and $m_{gas}$ refer to the mass of residuals and mass of emitted volatile gas, and it is assumed that they sum up to unity based on mass conservation. Although the loss of the mass of emitted volatile gas is redundant with the stoichiometric coefficients balancing, we found that it helps stabilize the training.

Since the model is stiff, we adopt the ODE integrator of Rosenbrock23 provided in the Julia package of DifferentialEquations.jl [38]. To facilitate the logarithmic operation in Eq. **3**, we have to clip the mass fraction to a non-negative value. We set $Y$ as $\max(eps, Y)$ and $eps = 1e - 8$ throughout the work. This clipping has the following effect on the reaction rate equation. Traditionally, the rate of reaction is calculated as in Eq. **2**. The clipping modifies that rate equation to the following:

$$r = [Cellu]^{v'_1}[S_2]^{v'_2}[S_3]^0 \phi([O_2]) AT^b \exp(-E_a/RT) \tag{8}$$

Where $\phi(O_2)$ is given by Eq. **9**.

$$\phi([O_2]) = \begin{cases} eps^{v'_{O_2}}, for\ [O_2] = 0 \\ [O_2]^{v'_{O_2}}, for\ [O_2] > 0 \end{cases} \tag{9}$$

This makes $v'_{O_2}$ a hyperparameter that defines the strength of an oxidation reaction, as will be discussed later.

Meanwhile, we set the absolute tolerance of the ODE solver equal to *eps*, as we found that it provides good numerical stability. One may use a smaller *eps* but this results in a higher computational cost due to a tighter tolerance for the ODE solver. It is worth noting that the choice of *eps* might affect the learned network as it introduces artificial intermediates with tiny concentrations during the induction period of pyrolysis. The recently developed differential programming package of DifferentialEquations.jl has readily enabled the computation of the gradients of the above loss functions to the CRNN parameters via differentiation over the ODE integrators. We employ the auto-differentiation framework of ForwardDiff.jl [39], which is more efficient than interpolation-based adjoint methods for stiff systems with the number of parameters in the orders of hundreds. Then, we can use the stochastic gradient descent optimization approach to learn the CRNN parameters, such as the popular optimizer of Adam proposed by Kingma and Ba [40] and implemented in the package of Flux.jl [41].

We treat the number of species and reactions as hyper-parameters of the CRNN model to be determined, which corresponds to the number of nodes in the input and output layer and the number of nodes in the hidden layer. While the optimal number of nodes could be different for different fuels and datasets, one can apply various hyper-parameter tuning approaches to automatically determine the optimal number of nodes. In this work, a grid searching approach is employed to determine the two hyper-parameters, i.e., increasing the number of proposed species and reactions until the model fitness cannot be further improved. It is noted that we shall weigh more of the number of species over the number of reactions if the goal is to build a model for large-scale fire simulation, since the number of species usually has a more significant impact on the computational cost of reacting flow simulations than that of the reactions.

Finally, we employ hard threshold pruning to further encourage sparsity in the learned CRNN weights, especially in the reaction orders and stoichiometric coefficients. The pruning proceeds by clipping the reaction orders and stoichiometric coefficients below a certain threshold. The threshold is also determined by grid searching as illustrated in [36]. Then, we can interpret and translate the pruned sparse CRNN model into the classical form of reaction equations.

### 3.2. Selection of Training and Test Data

We focus on using thermogravimetric data, since these data are available in the literature and the TGA is a standardized method of analyzing solids. Despite the significant interest in cellulose over the years, there are significant variations in the experimental data [42–44] due to the difference in the type of cellulose and experimental procedure used. We, therefore, restricted ourselves to the database of cellulose compiled by Richter et al. [33,45], who collected reported data that used very similar cellulose and experimental methodologies. The experimental results should therefore be consistent with each other. The adopted datasets compiled from four different references in [45] are listed in Table **1**. Nine of the 14 experiments are from pyrolysis in inert gas while the rest of the five are in an oxygen environment. Ten of them are under non-isothermal conditions with a constant heating rate while the rest of the four are under isothermal conditions. These experiments were also shown to be limited by kinetics and free of heat and mass transfer affects according to the threshold provided by Richter & Rein [46]. We divide the 14 experiments into training and test datasets, with the experiments of No. 2, 6, 9, 12 as the test datasets, which cover all of the references.

**Table 1**. Complete list of experiments with conditions: initial mass $m_0$, heating rate $\beta$ for non-isothermal (constant heating rate) conditions, temperature for isothermal conditions $T_s$ (constant furnace temperature).

| Exp | Reference | Group | $m_0$ (mg) | $\beta$ (°K/min) | $T_s$ (°K) | $[O_2]$ (%) |
|---|---|---|---|---|---|---|
| 1 | Varhegyi [47] | Training | 0.5 | 2 | - | 0 |
| 2 | Varhegyi [47] | Test | 0.5 | 10 | - | 0 |
| 3 | Varhegyi [47] | Training | 0.5 | 50 | - | 0 |
| 4 | Varhegyi [48] | Training | 3 | - | 523 | 0 |
| 5 | Varhegyi [48] | Training | 3 | - | 533 | 0 |
| 6 | Varhegyi [48] | Test | 3 | - | 548 | 0 |
| 7 | Varhegyi [48] | Training | 3 | - | 559 | 0 |
| 8 | Antal [44] | Training | 0.3 | 1 | - | 0 |
| 9 | Antal [44] | Test | 0.3 | 10 | - | 0 |
| 10 | Antal [44] | Training | 0.3 | 65 | - | 0 |
| 11 | Kashiwagi [49] | Training | 5 | 1 | - | 21 |
| 12 | Kashiwagi [49] | Test | 5 | 3 | - | 21 |
| 13 | Kashiwagi [49] | Training | 5 | 5 | - | 21 |
| 14 | Kashiwagi [49] | Training | 5 | 3 | - | 5.2 |

## 3.3. Training Strategies and Weight Pruning

In general, there are no universal rules on the choice of hyper-parameters, such as learning rate, for the training of a neural network model. Here, we present the learning strategies that are adopted in this work and can efficiently learn the cellulose kinetic modes. Figure **3** shows the typical history of the loss functions and the $L_2$ norm of gradients. The model is trained for 2500 epochs and each epoch consists of ten times of parameter updates by iterate over ten experimental conditions. As seen from the history of the loss function in Fig. **3a**, the loss on the training dataset converges after 500 epochs. We let the training run another 2000 epochs as the test loss is still decreasing. In addition, we use the annealing of learning rate to fine tune the model. The initial learning rate of 5e-3 and the learning rate decays by a factor of 0.2 every 500 epochs until the minimum of 1e-5 is reached. We also apply gradient clipping with a threshold of 1e2 since we observed abnormal gradient explosion during the training as shown in Fig. **3b**. The neural ODEs can be viewed as very deep residual neural networks with each time-step corresponding to a block of a residual network [37]. Gradient explosion is a common issue in very deep neural networks [50], and gradient clipping is a widely adopted technique to partially mitigate the issue of gradient explosion. The gradient clipping proceeds by multiplying the gradients of model parameters by a factor smaller than unity, such that the norm of the gradients is equal to the pre-defined threshold of 1e2 when the norm of the gradients exceeds the threshold.

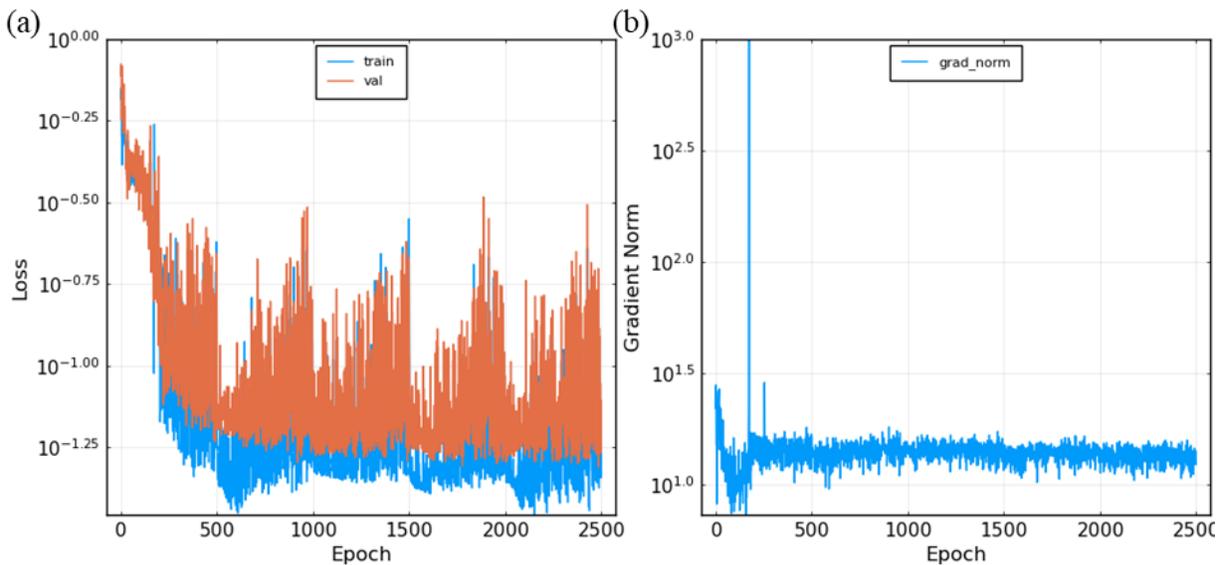

**Figure 3.** (a) The typical evolution of loss functions and (b) the $L_2$ norm of gradients with the number of epochs. Results shown correspond to the CRNN with four species and six reactions. The maximum gradient norm is 1e60 and we clip it for better visualization.

We then present the learned CRNN weights in Table. **2a**. The dimensions of inputs and hidden nodes are physically interpretable as each row of the matrix corresponds to a reaction. The weights in the input layer reveal the reaction orders of each reaction, and they are always positive. The weights in the output layer correspond to the stoichiometric coefficients, and positive/negative values indicate that the corresponding species are produced/consumed, whereas zero values indicate that the corresponding species do not participate in the reaction. The biases reflect the rate constants. Consequently, we can infer the corresponding eight reaction pathways from the learned CRNN weights.

There are relatively small (compared to unity) nonzero input and output weights in the learned CRNN model, and we apply hard-threshold pruning to the reaction orders and stoichiometric coefficients to produce a sparse model and better interpretation. The threshold is also determined via grid searching, which means it does not have to be defined by the user Figure **4** shows the dependence of the loss function with the threshold value. It is found that there are sharp changes in fitness when the threshold is larger than 0.185, and thus 0.185 is chosen as the threshold. Table **2b** shows the learned CRNN weights after pruning. Noting that the reactions R2 and R8 in Table **2a** are omitted after pruning since the stoichiometric coefficients and reaction orders are pruned to be zero.

**Table 2**. Learned CRNN weights and biases with four species and six reactions: (a) before pruning, and (b) after pruning with the threshold of 0.185.

(a) Learned CRNN weights before pruning.

| Cellu | S2 | S3 | O2 | $E_a$ | b | lnA | Cellu | S2 | S3 | Vola |
|---|---|---|---|---|---|---|---|---|---|---|
| 0 | 0 | 1.32 | 0.45 | 88.7 | 0.05 | 22.69 | 0 | 0.56 | -1.32 | 0.76 |
| 0 | 0 | 0.05 | 0.17 | 228.6 | 0 | 16.91 | 0 | 0.05 | -0.05 | 0 |
| 0.2 | 0.04 | 0 | 0.06 | 222.4 | 0 | 17.3 | -0.2 | -0.04 | 0.21 | 0.04 |
| 1.52 | 0 | 0 | 0.19 | 187.4 | 0.04 | 33.64 | -1.52 | 0.38 | 0.46 | 0.68 |
| 0.4 | 0 | 0.61 | 0.09 | 117.2 | 0.15 | 14.92 | -0.4 | 0.41 | -0.61 | 0.61 |
| 1.15 | 0.38 | 0 | 0.01 | 218.0 | 0.34 | 36.75 | -1.15 | -0.38 | 0.64 | 0.89 |
| 0 | 1.91 | 0 | 0.33 | 110.5 | 0.03 | 14.08 | 0 | -1.91 | 1.27 | 0.63 |
| 0 | 0.05 | 0 | 0.16 | 227.3 | 0 | 17.14 | 0 | -0.05 | 0.05 | 0 |

(b) Learned CRNN weights after pruning with the threshold of 0.185.

| Cellu | S2 | S3 | O2 | $E_a$ | b | lnA | Cellu | S2 | S3 | Vola |
|---|---|---|---|---|---|---|---|---|---|---|
| 0 | 0 | 1.32 | 0.45 | 88.7 | 0.05 | 22.69 | 0 | 0.56 | -1.32 | 0.76 |
| 0.2 | 0 | 0 | 0 | 222.4 | 0 | 17.3 | -0.2 | 0 | 0.2 | 0 |
| 1.52 | 0 | 0 | 0.19 | 187.4 | 0.04 | 33.64 | -1.52 | 0.38 | 0.46 | 0.68 |
| 0.4 | 0 | 0.61 | 0 | 117.2 | 0.15 | 14.92 | -0.4 | 0.41 | -0.61 | 0.61 |
| 1.15 | 0.38 | 0 | 0 | 218.0 | 0.34 | 36.75 | -1.15 | -0.38 | 0.64 | 0.89 |
| 0 | 1.91 | 0 | 0.33 | 110.5 | 0.03 | 14.08 | 0 | -1.91 | 1.27 | 0.63 |

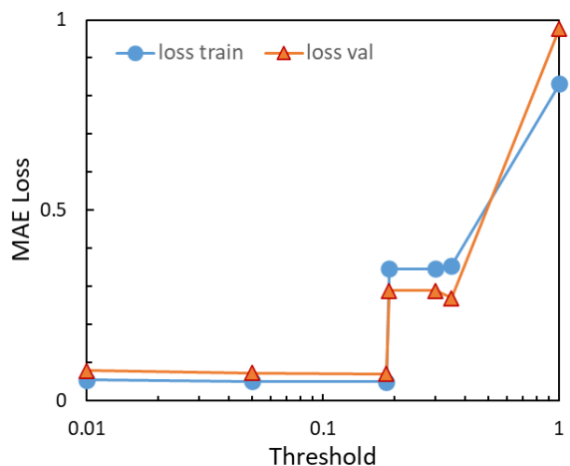

**Figure 4.** Dependence of minimum loss functions for the training and test datasets on the pruning threshold. There is a sudden jump at the threshold value of 0.185 and thus 0.185 is chosen to prune the CRNN model in this work.

## 4. Results

### 4.1. The plausibility, validity, and interpretability of the derived kinetic model for cellulose

In this section, we will discuss the derived kinetic model by applying the CRNN method [36] to the pyrolysis and oxidation of cellulose. To this end, we build a neural network (CRNN) that takes three chemical species as an input (cellulose, S2, S3) as well as oxygen. The choice of three input species is justified in Section 4.2. The code automatically finds all reactions between the three species and oxygen as well as the corresponding kinetic parameters for rate constants in Arrhenius form. We trained the model for 2500 epochs on 14 experiments of cellulose. Afterwards, weight pruning was applied to find the optimal kinetic model.

Our method proposes a plausible kinetic model of four species (cellulose, 2 product species, and volatiles) and six reactions, which is well within the range of current optimal reduced kinetic models (5 species, 5 reactions) [51]. The reaction scheme is depicted in Fig. **5** and the kinetic parameters are listed in Table **3**. Figure **6** then shows the comparisons between the predictions and experimental data. Overall, the predictions and measurements agree well across all of the experimental datasets. We found a mean absolute error of 0.03 across all experiments (0.037 in pyrolysis, 0.014 in oxidation conditions) which is similar to the errors (>0.06) of the current well-established literature models [33]. Notably, the errors in Richter et al. [33] were derived for a larger set of experiments so that it is reasonable that our model shows a lower average error than reported in [33]. They do span a range of optimized and literature models, which should make them representative of the literature. A re-analysis of Richter et al. [33] data is expected to yield similar results to those obtained here. In addition, the measured and predicted mass loss rate is presented in the Figure **S1** of Supplemental Material, which also shows good agreement between the measurements and predictions. Based on these results, we can conclude that CRNN derives plausible and accurate reaction schemes.

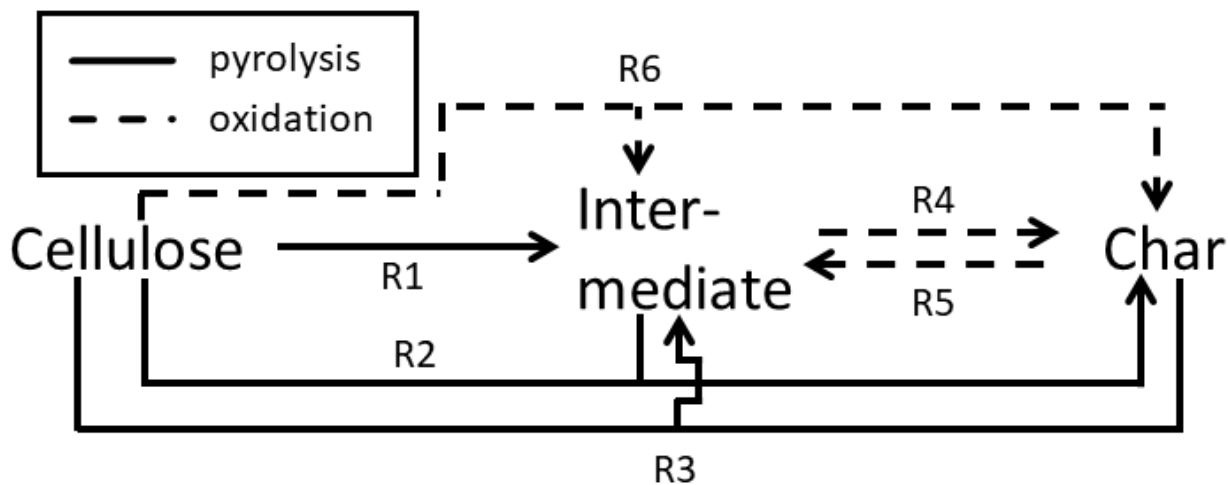

**Figure 5.** Learned reaction scheme that corresponds to the reactions in Table 3.

**Table 3**. Learned Pathways Interpreted from the CRNN.

| | Pathways | $E_a$ (kcal/mol) | $b$ | $lnA$ |
|---|---|---|---|---|
| R1 | $0.2\ Cellu \rightarrow 0.2\ S3$ | 222.4 | 0 | 17.3 |
| R2 | $0.4\ Cellu + 0.61\ S3 \rightarrow 0.41\ S2 + 0.61\ Vola$ | 117.2 | 0.15 | 14.92 |
| R3 | $1.15\ Cellu + 0.38\ S2 \rightarrow 0.64\ S3 + 0.89\ Vola$ | 218.0 | 0.34 | 36.75 |
| R4 | $1.32\ S3 + 0.45\ O2 \rightarrow 0.56\ S2 + 0.76\ Vola$ | 88.7 | 0.05 | 22.69 |
| R5 | $1.91\ S2 + 0.33\ O2 \rightarrow 1.27\ S3 + 0.63\ Vola$ | 110.5 | 0.03 | 14.08 |
| R6 | $1.52\ Cellu + 0.19\ O2 \rightarrow 0.38\ S2 + 0.46\ S3 + 0.68\ Vola$ | 187.4 | 0.04 | 33.64 |

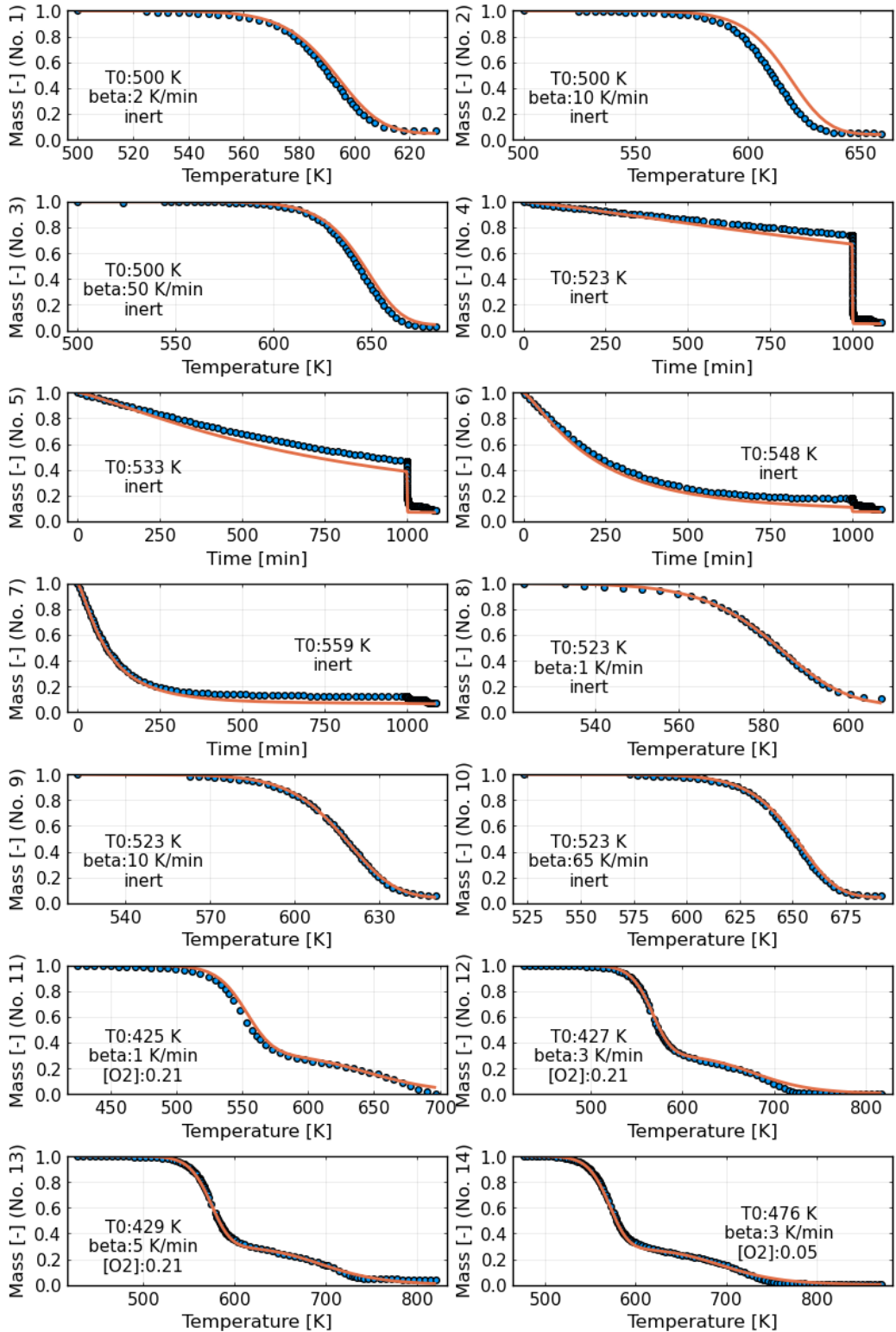

**Figure 6.** Measured residual mass profiles (symbols) and the predictions (solid lines) using the learned CRNN model presented in Table 3.

The physical interpretation of the derived chemical species also agrees well with the current literature. Our CRNN model is in the same form as conventional kinetic models, which allows us to employ various standard analysis tools [52] to understand the chemical insights of the pyrolysis process. Note that one peculiarity is the treatment of oxidation reactions in our model. Our set-up allows a reaction to simultaneously be a pyrolysis and oxidation reaction, with their respective strength being determined by the hyper parameter $v_{O_2}$. A value of 0 indicates a pure pyrolysis reaction, and a value of 1 or higher indicates a pure oxidation reaction in almost all cases. A value between 0 or 1 indicates that oxygen acts as a catalyst for the reaction pathway and enhances the speed of reaction. In other words, we use the oxygen concentration to adjust the pre-exponential factor. With this in mind, the species profile in Fig. **7** allows us to better understand the roles of S2 and S3. Based on the reaction pathways in Table **3**, we can conclude that the reaction is initiated with R6, which is much faster than R1. Based on the predicted species profiles in Fig. **7**, the production of species S3 occurs earlier than that of S2; however, the concentration of S3 is negligible at the end of the experiment. On the other hand, the concentration of S2 has a concentration of about 0 to 7% on an initial mass basis. Although Fig. **7** only shows the predictions corresponding to the experimental conditions of No. 2, similar trends in the evolutions of S2 and S3 have been found in all cases. This suggests that S3 is the species that initiate the reaction process, while S2 is the main solid product. We could therefore interpret S3 as either active cellulose, tar, or a combination of them. In other words, S3 can be interpretated as an intermediate as its concentration is zero at the beginning and end of the experiment. On the other hand, S2 can be interpreted as a solid char. As a result, we can conclude that the species found by the CRNN method agree with the literature. We will hence forth call S3 the intermediate species and S2 the char species.

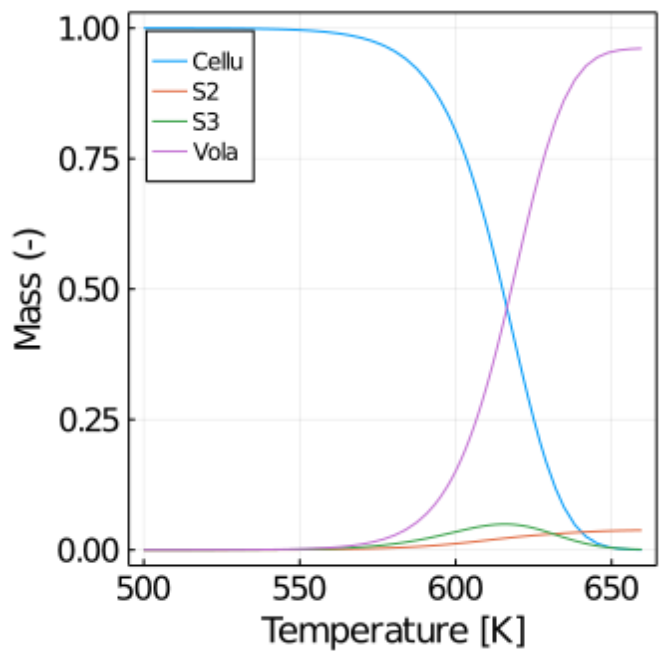

**Figure 7.** Species profiles for experimental conditions of No. 2 predicted using the learned CRNN model presented in Table 3.

Previous studies identified five key chemical observations for cellulose, and the reaction scheme derived from the CRNN reveals similar insights.

**(I)** The reaction scheme should have competing pyrolysis reactions to be able to reduce the variation in char yield and heat of pyrolysis with heating rate [53,54]. Our reaction scheme shows two competing reactions as shown in Fig. **5**.

**(II)** The reaction scheme must have at least one pathway that leads to a negligible char yield. This pathway accounts for the observation by Dauenhauer's group [55] and others [56] that the char yield of cellulose becomes or approaches zero under rapid heating conditions. This could be achieved in our cases through R1 followed by a back and forth between S2 (char) and the S3 (intermediate) with R2 and R3 or R4 and R5.

**(III)** The reaction scheme must have an intermediate to char. This intermediate was first speculated by Broido [57] who observed a change in char yield when a sample was preheated at low temperature before the experiments. Anca-Couce [34] provided a review on the current evidence on the intermediate, and Dauenhauer [55] provides visual evidence for it. Based on his

- literature review, Anca-Couce speculated that the majority or even all the char is formed through an intermediate. His hypothesis is consistent with our reactions (R2 & R3 and R4 & R5) between S2 (char) and S3 (intermediate), who propose that the intermediate (S2) and char (S3) can form each other.

**(IV)** The reaction scheme should allow oxygen to accelerate the decomposition of cellulose to char. Mamleev [58] showed that oxygen has a catalytic effect on the decomposition of cellulose. Previous studies [45,49] account for this phenomena through competing reactions, and our methodology arrived at the same result. Our orders of oxidation for R4 & R6 are between 0 and 1, which represents in our mathematical formulation a catalytic effect by oxygen.

**(V)** Oxygen should be able to oxidize char. The burn-off of char by oxygen is a well-established combustion phenomenon [59]. Our methodology independently found this reaction in R5, but with the limitation that our model did not find inorganic content as a separate species.

We can therefore conclude that our reaction scheme agrees with the literature as it satisfies all major chemical observation for cellulose. In addition, we found that kinetic parameters agree well with those of the literature. We will restrict ourselves to the discussion of the kinetic parameters of the pyrolysis reactions as high-fidelity and atomistic calculations exist for these. For the oxidation of the only criteria that the oxidation reactions should be faster than the pyrolysis reactions based on the experiments of [49],this criterion is fulfilled.  We previously established that R1 and R6 are the initiation reaction of the whole process . R2 and R4 are the char producing reactions. Both of these reactions map well on the currently leading models for cellulose (Broido-Shafizadeh Scheme). This would leave R3 as the main volatile producing reaction in the Broido-Shafizadeh scheme. However, the mapping breaks down at this point as the main volatile reaction in the traditional models has the intermediate as a reactant and not a product. We will, therefore, treat the R2 & R3 and R4 & R5 jointly as the volatile reactions. This step is necessary as previous models have not explicitly considered char reacting to a different solid. Figure **8** shows the comparison of all pure pyrolysis reactions obtained by CRNN with data from the literature. The shaded area represents the currently agreed upon range of rate constant for the production of volatiles from cellulose. The values are derived from TGA experiments and atomistic calculations [35,60] by assuming a one-step reaction scheme. It is expected that the main reaction producing volatiles follows within or close to this range. Figure **8** shows that this is the case for our kinetic model, as the CRNN model found that all volatile producing reactions are either within or close to that range. Also, Krumm et al. [61] measured in high-fidelity experiments the decay of a surrogate of cellulose as well as the production of volatiles which is also indicated in Fig. **8**. We can see that their measurements agree well with R5, as would be expected from a good model. These measurements, therefore, provide evidence for the accuracy of our kinetic parameters.

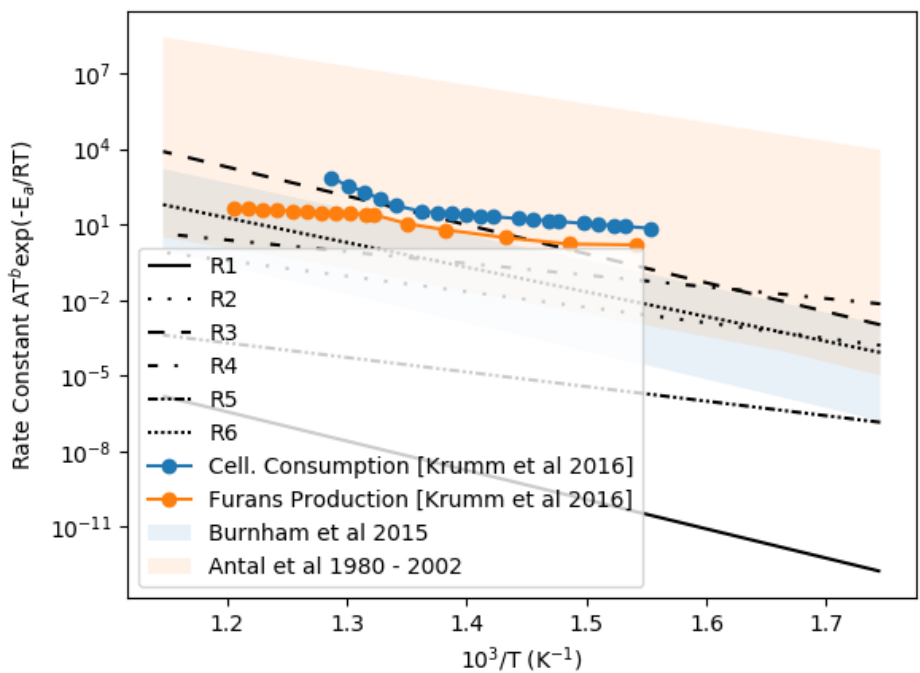

**Figure 8.** Comparisons of the determined rate constants for the pyrolysis reactions with the state-of-the-art experiments as well as the range determined from atomistic calculations for the production of volatiles (shaded areas).

In summary, the CRNN method derived a plausible reaction scheme with plausible kinetic parameters. The reaction scheme agrees with all main chemical observations. The kinetic parameters are close to the currently agreed optimal values. This evidence shows that the CRNN methods can derive practically useful and scientifically sound kinetic models for biomass from a small number of TGA measurements. Our experience from this paper suggests that 3 isothermal and 3 non-isothermal experiments would be sufficient to derive a good kinetic model within one day. We believe this to be a remarkable feat as the observation and values we compare to have been derived over several decades. In the next section, we will therefore discuss potential extensions and applications of our methodology.

### 4.2. Sensitivities of the Number of Proposed Species and Reactions

Finally, we briefly discuss the sensitivities of the model performance to the number of proposed species and reactions. While the optimal number of nodes could be different for different fuels and datasets, hyper-parameter tuning approaches are applied here to automatically determine the optimal number of nodes (i.e., the number of reactants, intermediate species, and products). Figure **9** shows the dependence of minimum loss functions for the test datasets on the number of proposed species and reactions. The hyper-parameter exploration takes about 100 CPU hours. Lower loss functions, shown in darker color, correspond to better fitness. To facilitate parametric study, we apply the same training procedure to all of the models. Overall, the fitness is improved with larger models, i.e., more species and nodes, although exceptions occur. These exceptions can be attributed to the bias-variance tradeoff in machine learning. We can then choose the model size according to the Occam razor theory and our expectation of the model size. The Occam razor theory suggests choosing a model as simple as possible but with an acceptable prediction capability. Since the model size has a substantial impact on the integration into large-scale fire simulation codes, we could also consider the trade-off between model size and model accuracy [32,62]. Therefore, we chose the size of four species and eight reactions in this work as they provide reasonably good prediction capability and visually better performance than three species or fewer reactions. Note that while the eight reactions are reduced to six after pruning, it might be difficult to reach the same performance if we initially proposed a six reaction model. While the optimization of neural networks is still an open question, there is some recent theoretical analysis [63] that suggests that over-parameterized models are more likely to overcome the local minima with SGD optimization and achieves better model performance. Furthermore, with the help of the weight pruning in identifying the unimportant reactions and species, the eventual models will be less sensitive to the choice of the initially proposed model size.

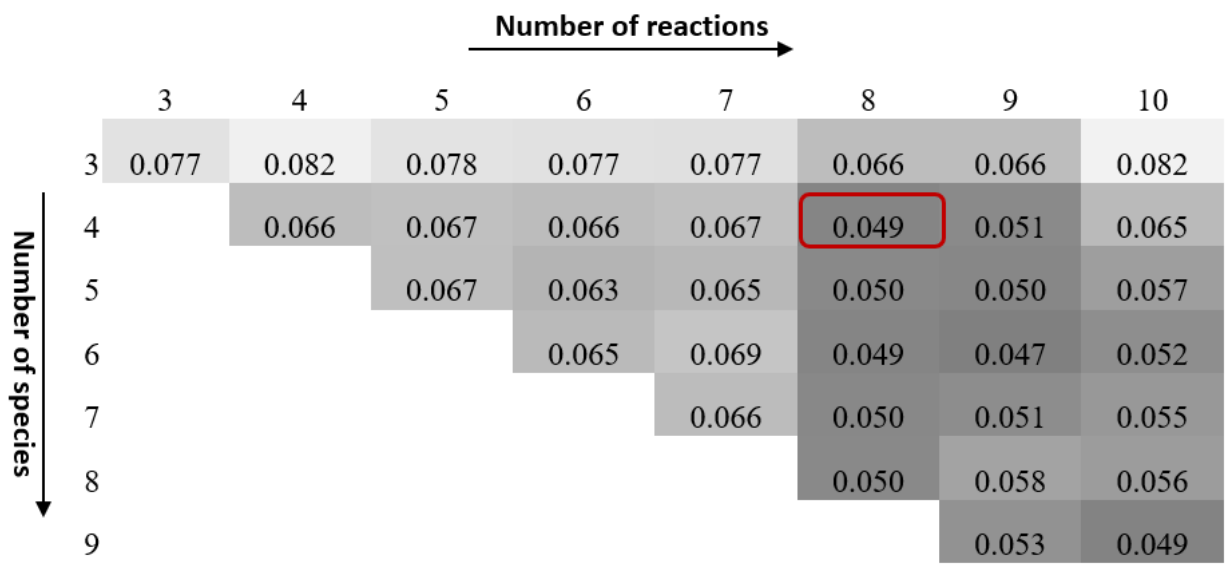

**Figure 9.** Dependence of minimum loss functions for the test datasets on the number of proposed species and reactions. Colors are rendered according to the values of the loss functions. Lower loss functions, shown as in darker color, correspond to better fitness. Boxed is the one chosen in this work.

### 4.3. Practical and Future Applications of the Proposed Methodology

The combination of neural network models and physics-based models can also bridge the model compression techniques developed in machine learning communities and the mechanism reduction tools developed in combustion communities. This work has presented the application of weight pruning developed by machine learning communities for model reduction. After the learned model is translated into the conventional form of the reaction mechanism, we can readily apply mechanism reduction methods such as directed relation graph (DRG) [64,65] and path flux analysis [66] to further reduce the model. Since it is not straightforward to apply weight pruning to the non-linear Arrhenius form of rate constants, path flux-based methods such as DRG could be superior. In another example, the differential programming can be further coupled with a Bayesian inference to efficiently characterize the uncertainty in the learned kinetic model and guide design of future experiments.

While the current studies focused on the pyrolysis in the condensed phase, we can further couple the CRNN model with gas-phase kinetic and thermodynamic models to learn the pyrolysis sub-models from combustion experiments of biomass, e.g., the burning rate and emissions. Since gas-phase kinetic models are relatively well studied compared to the condensed phase reactions, we can rely on existing models instead of learning them from experimental data. Recently developed differential combustion simulation packages, like Arrhenius.jl [67], could seamlessly integrate existing gas-phase kinetic models with CRNN models to predict combustion behaviors and backpropagate the error to the pyrolysis sub-models. In addition, with the development of physics-informed neural networks [68], one can also learn the pyrolysis sub-models from experimental data with flow-chemistry interactions characterized by partial differential equations.

While cellulose kinetic models are relatively well understood compared to live fuels involved in wildland fires and new types of construction materials, we expect that the success of modeling cellulose pyrolysis without prior knowledge and the capacity of neural network models will substantially accelerate our modeling and understanding of those real-world complex fuels.

## 5. Conclusions

This work presents a machine learning framework based on Chemical Reaction Neural Networks for autonomously modeling the kinetics of biomass pyrolysis. The capability of simultaneously inferring reaction pathways, intermediate species, and kinetic parameters from TGA data is demonstrated in the pyrolysis modeling of cellulose. The learned kinetic model is not only able to accurately predict the residual mass profiles but also facilitate revealing chemical insights using various kinetic analysis tools. The inference of unknown intermediate species can also guide the design of experimental diagnostics to identify those potential intermediate species.

While this work focused on learning interpretable neural network models for kinetic sub-models, the framework can be extended to learn other sub-models as well, such as thermodynamic models or diffusion sub-models, by incorporating relevant physics laws and formulas into the neural network architectures.

## 6. Acknowledgement

SD would like to acknowledge the support from the Karl Chang (1965) Innovation Fund at Massachusetts Institute of Technology. FR and MJG were supported by the National Institute of Standards and Technology Disaster Resilience grant 70NANB19H053. WJ would like to thank the fruitful discussions with Professor Guillermo Rein at Imperial College London.

## 7. Nomenclature

$b$: temperature dependence coefficient for the pre-exponential factor

*Cellu*: Cellulose

CRNN: chemical reaction neural networks

$E_a$: activation energy for the reaction

ln A: pre-exponential factor in logarithmic scale

Loss: loss function for optimization

MAE: mean absolute error

$S_i$: intermediate species

TGA: thermogravimetric analyzer

*Vola*: volatile

## Supplemental Materials

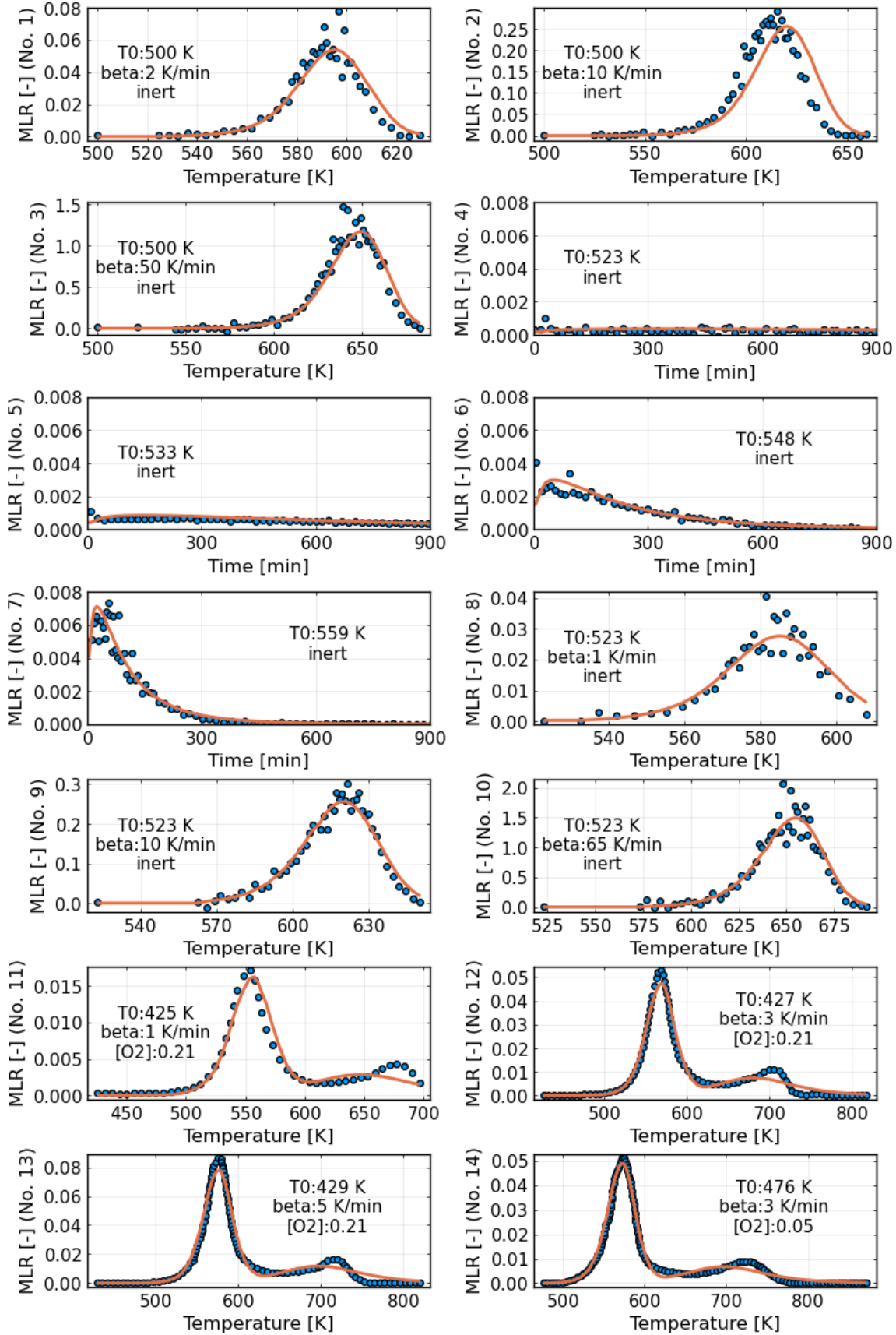

Figure S1. Measured mass loss rate (MLR) profiles (symbols) and the predictions (solid lines) using the learned CRNN model.